\documentclass[10pt, a4paper]{article}
\usepackage[utf8]{inputenc}
\usepackage{amsmath}
\usepackage{amssymb}
\usepackage{enumerate}
\usepackage{enumitem}
\usepackage{graphicx}
\usepackage{siunitx}
\usepackage{float}
\usepackage{dirtytalk}

\usepackage{bm}% bold math

\usepackage[font=footnotesize]{caption}

\usepackage{titlesec}
\titleformat{\section}
{\bf\sffamily}
{\thesection. }
{5pt}
{\MakeUppercase}
\renewcommand{\thesection}{\Roman{section}} 

\titleformat{\subsection}
{\sffamily}
{\thesubsection. }
{5pt}{}

\renewcommand{\thesubsection}{\Alph{subsection}}

\renewcommand{\thesubsection}{\Alph{subsection}}

\titleformat{\subsubsection}
{\sffamily}
{\thesubsubsection. }
{5pt}{}

\usepackage[affil-it]{authblk}

\usepackage{abstract}
\usepackage[hidelinks]{hyperref}
\usepackage{xcolor}
\hypersetup{
	colorlinks,
	linkcolor={blue!50!black},%{red!80!black},
	citecolor={blue!50!black},
	urlcolor={blue!80!black}
}

\usepackage[citestyle=numeric,style=phys,biblabel=brackets,backend=bibtex,sorting=none,url=false,doi=false, natbib]{biblatex}
\bibliography{bibliography}
\DefineBibliographyStrings{english}{andothers={\itshape et\addabbrvspace al\adddot}}

\usepackage[lof,lot]{etoc}

\usepackage{geometry} % Dimension of body text
\usepackage[]{parskip} %Insert after packages for references %skip=0.7em, indent=1.5em

\usepackage[left,mathlines]{lineno}

\title{On Linear and Non-Linear Mechanics of Cyanobacterial Colonies}
\author[1]{Yuri Z. Sinzato\footnote{y.z.sinzato@uva.nl}}
\author[2,3]{Annemieke M. Drost}
\author[2,3]{Dedmer B. Van de Waal}
\author[4]{Robert Uittenbogaard}
\author[3]{Petra M. Visser}
\author[3]{Jef Huisman}
\author[1,5]{Maziyar Jalaal\footnote{m.jalaal@uva.nl, m.jalaal@damtp.cam.ac.uk}}

\affil[1]{Van der Waals-Zeeman Institute, Institute of Physics, \protect\\
University of Amsterdam, Science Park 904, Amsterdam, 1098XH, The Netherlands}
\affil[2]{Department of Aquatic Ecology, Netherlands Institute of Ecology (NIOO-KNAW),  \protect\\Wageningen, 6708 PB, The Netherlands}
\affil[3]{Department of Freshwater and Marine Ecology, Institute for Biodiversity and Ecosystem Dynamics, \protect\\
University of Amsterdam, Science Park 904, Amsterdam, 1098XH, The Netherlands}
\affil[4]{Hydro-Key Ltd, Haelen 6081EA, The Netherlands}
\affil[5]{Department of Applied Mathematics and Theoretical Physics, \protect\\
University of Cambridge, Wilberforce Road, Cambridge CB3 0WA, United Kingdom}

\begin{document}
\begingroup
\sffamily
\date{}
\maketitle
\endgroup

\begin{abstract}

Toxic cyanobacterial blooms are a growing environmental concern that affects freshwater ecosystems, drinking water supplies, and public health. The cyanobacterium \textit{Microcystis} is among the most important bloom forming species. It often grows in large colonies, which enhances its flotation, reduces grazing, and improves nutrient regulation. \textit{Microcystis} cells are held together by a matrix of extracellular polymeric substances (EPS), making colony mechanics crucial for bloom formation. However, an analysis of the biomechanical properties of cyanobacterial colonies, and how these properties relate to environmental conditions like nutrient availability, remains largely missing. Here, we use micropipette force sensors to quantify the linear and non-linear mechanical properties of individual colonies at single-cell resolution.  Bulk shear rheology complements  these measurements by probing macroscopic properties. The measured tensile strength and yield stress are broadly comparable to those of bacterial biofilms and are far greater than the hydrodynamic stresses typically found in wind-mixed lakes. This implies that cyanobacterial colonies are highly resistant to fragmentation by natural mixing processes. We also show that low nutrient availability, particularly low phosphorus, produced stronger colonies, suggesting structural changes in the EPS. Overall, our results establish mechanical testing as a tool for a more complete and physically grounded understanding of cyanobacterial colony formation. 

\end{abstract}

%%%%%%%%%% 
% MAIN  TEXT
%%%%%%%%%%%

% Start reference section
\begin{refsection}

\section{Introduction}

Cyanobacterial blooms have become an increasing global concern in freshwater and brackish aquatic systems, with profound consequences for ecosystem health, biodiversity, and drinking water security \cite{harke2016review,huisman2018cyanobacterial,griffith_harmful_2020,chorus_toxic_2021}. Among the bloom-forming species, the cosmopolitan cyanobacterium \textit{Microcystis} is particularly problematic, as it frequently forms dense and toxic surface accumulations that disrupt aquatic food webs and limit access to safe water supplies. A defining feature underlying the ecological success of \textit{Microcystis} is its ability to form colonies, a trait that confers multiple advantages, including enhanced protection against grazing, increased flotation velocity, and improved regulation of nutrient uptake \cite{xiao_colony_2018,huisman_changes_2004}. Much like bacterial biofilms, the persistence of these colonial structures relies on withstanding fluctuating environmental stresses, among which are mechanical forces \cite{stewart_biophysics_2014,Sinzato_2026}. Understanding the biomechanics of cyanobacterial colonies is therefore crucial to gain deeper insight into how blooms form, persist, and dominate in dynamic aquatic environments.

The cells of \textit{Microcystis} colonies are held together by a matrix of bound extracellular polymeric substances (EPS), composed primarily of secreted polysaccharides \cite{le_microcystis_2022}. This bound EPS can be further subdivided into loosely bound and tightly bound fractions, classified according to their resistance to removal \cite{xu_investigation_2013}. The production and composition of EPS in \textit{Microcystis} colonies are regulated by biotic and abiotic factors, among which nutrient availability plays a central role \cite{xiao_colony_2018}. In particular, nitrogen (N) and phosphorus (P) exert a dual influence on colony formation: high nutrient concentrations enhance cellular growth rates, allowing colonies to increase in size through cell division \cite{shen_comparative_2007,duan_colony_2018}, whereas nutrient limitation promotes the allocation of fixed carbon into bound EPS, which is likewise essential for maintaining colonial cohesion \cite{duan_ecological_2021,wang_effect_2011,feng_role_2022}. This dual role is often observed during bloom development, when dissolved nutrients are abundant during early stages but may become limiting as biomass accumulates \cite{paerl_nutrient_2015}. Consistent with this picture, sustained high nutrient availability in culture media has been suggested as a key factor underlying the frequent loss of colonial morphologies in laboratory strains of cyanobacteria \cite{reynolds_variability_2007,xiao_colony_2018}.

Beyond its biochemical and physiological roles, the EPS matrix directly determines the mechanical integrity of cyanobacterial colonies, which is a critical trait for their ecological success. \textit{Microcystis} colonies can withstand fragmentation under the hydrodynamic stresses typical of wind-mixed lakes \cite{li_effect_2013,obrien_disaggregation_2004}. Under sufficiently strong shear in controlled flow conditions, however, erosion of single cells from the colony periphery occurs, leading to reductions in colony size \cite{Sinzato_2026} and associated changes in colony morphology \cite{feng_structural_2020}. Strong cohesive forces may also serve as an effective defence mechanism against grazing, because high mechanical strength of the colony may prevent its fragmentation into ingestible pieces under the mechanical stresses imposed during feeding \cite{stewart_biophysics_2014,gerritsen_not_1988,van_wichelen_strong_2010}. In addition, a robust EPS layer can partially impede digestion by large filter-feeders, providing an additional protective advantage \cite{reynolds_variability_2007,dionisio_pires_grazing_2005}. Despite the central role of mechanical properties in the formation, persistence, and ecological impact of cyanobacterial blooms, direct measurements of the mechanical properties of cyanobacterial colonies remain scarce \cite{Sinzato_2026,vignaga_quantifying_2012}.

Mechanical testing methods have been developed for heterotrophic bacterial biofilms \cite{gordon_biofilms_2017,boudarel_towards_2018}, yet the reported material properties vary widely depending on species, measurement scale, and direction and mode of applied forcing. An experimental approach that can probe the mechanics of natural cyanobacterial colonies has not been described yet. Micropipette force sensing is a micromanipulation technique that infers forces from the controlled deflection of a flexible hollow glass pipette \cite{backholm_micropipette_2019}. Its single-cell spatial resolution makes it well suited for probing local mechanical properties within cyanobacterial colonies, while its minimal sample-volume requirement is advantageous for field samples, where cyanobacterial abundance may be low or mixed with other phytoplankton, rendering bulk mechanical tests impractical \cite{pavlovsky_effects_2015}. Micropipette-based force measurements have previously been used to quantify the mechanical strength of bacterial biofilms with size scales comparable to cyanobacterial colonies \cite{poppele_micro-cantilever_2003,aggarwal_development_2010,aydin_strong_2026}, as well as other (multi)cellular aggregates \cite{askari_soft_2025}. Complementary to microscale mechanical tests, bulk shear rheology provides a useful set of techniques that has been recently applied to identify macroscopic solid- and fluid-like responses in bacterial biofilms \cite{doi:10.1073/pnas.2512757123,pavlovsky_effects_2015} and laboratory colonies of \textit{Microcystis} \cite{Sinzato_2026}.

Here, we used micropipette force sensing to investigate the mechanical properties of field-collected \textit{Microcystis} spp. colonies. First, we quantified the force required to detach a single cell from a colony, from which we extracted the microscale tensile strength and Young’s modulus. We identified the colony region with the highest deformation by imaging the cells and the fluorescently labelled EPS layer. Secondly, we complemented these microscale measurements by bulk shear rheology on macroscopic aggregates of \textit{Microcystis} colonies, allowing a direct comparison between microscale cohesion and macroscopic mechanical responses. Finally, to investigate how colony mechanics and morphology respond to environmental conditions, we used microcosms  \cite{https://doi.org/10.1890/13-1251.1,https://doi.org/10.1111/gcb.14660} in which a natural cyanobacterial community dominated by \textit{Microcystis} was exposed to different nitrate and phosphate concentrations. While low nutrient levels were previously shown to enhance EPS excretion  \cite{duan_ecological_2021}, we investigated the hypothesis that an increase in proportion of EPS per colony results in a higher mechanical strength.

\section{Results and Discussion}

\subsection{Linear and non-linear mechanical properties of \textit{Microcystis} colonies}

\begin{figure}[t!]
    \centering
    \includegraphics[width = \columnwidth]{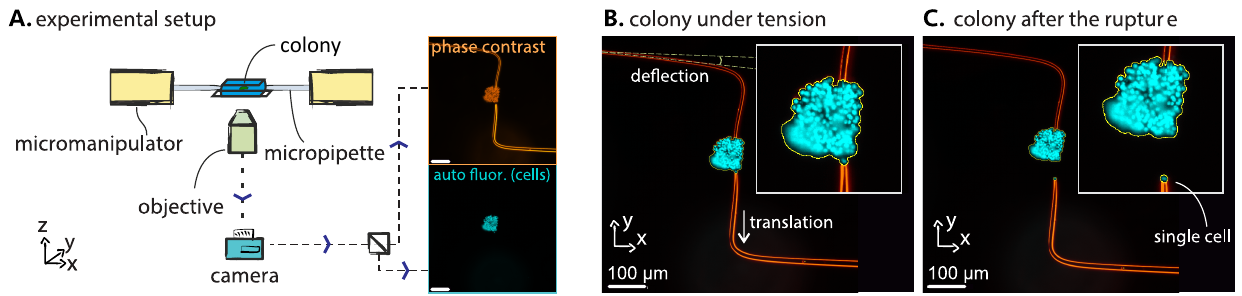}
    \caption{Experimental setup to measure the force required to remove a single cell from a \textit{Microcystis} colony. (A) A colony is held between two L-shaped micropipettes, each controlled by a 3-axis automated micromanipulator. The micropipettes (orange) are imaged using phase-contrast microscopy, while the cells (cyan) are imaged using wide-field fluorescence with excitation of chlorophyll \textit{a}. Scale bar = \qty{100}{\um}. (B) A rigid micropipette (bottom right) is translated at a constant speed while the deflection of a flexible micropipette (top left) is measured. (C) Upon rupture, a single cell is removed from the main colony, and the flexible micropipette returns to its equilibrium position. Insets show details of the colony, where the yellow line indicates the outline of the colony. See Supplementary Movie S1}
    \label{fig:setup}
\end{figure}

\textit{Microcystis} spp. colonies were mechanically characterized with micropipette force sensors (Fig. \ref{fig:setup}A, \nameref{sec:methods}).  Each colony was held between two L-shaped micropipettes and quasi-static tensile tests were performed (Supplementary Movie S1). The micropipettes and the cells within the held colony were visualized using phase-contrast microscopy and fluorescence microscopy, respectively. The rigid micropipette pulled the colony at a constant speed, while the tensile force $F$ applied to the colony was calculated from the deflection of the flexible micropipette (Fig. \ref{fig:setup}B, \nameref{sec:methods}). For each test, the extension was conducted until the fracture of the colony (Fig. \ref{fig:setup}C) and the maximum recorded force was defined as the fracture force $F_b$. The fragments were composed of a single cell in 91\% of the tests, while the remaining tests produced fragments of two or three cells. The large percentage of single fragments is consistent with our previous study, which identified that erosion of single cells is the primary fragmentation mechanism under strong flows \cite{Sinzato_2026}. For consistency in the comparison of forces, only the tensile tests with a single cell fragment were analysed.

\begin{figure}[t!]
    \centering
    \includegraphics[width=\textwidth]{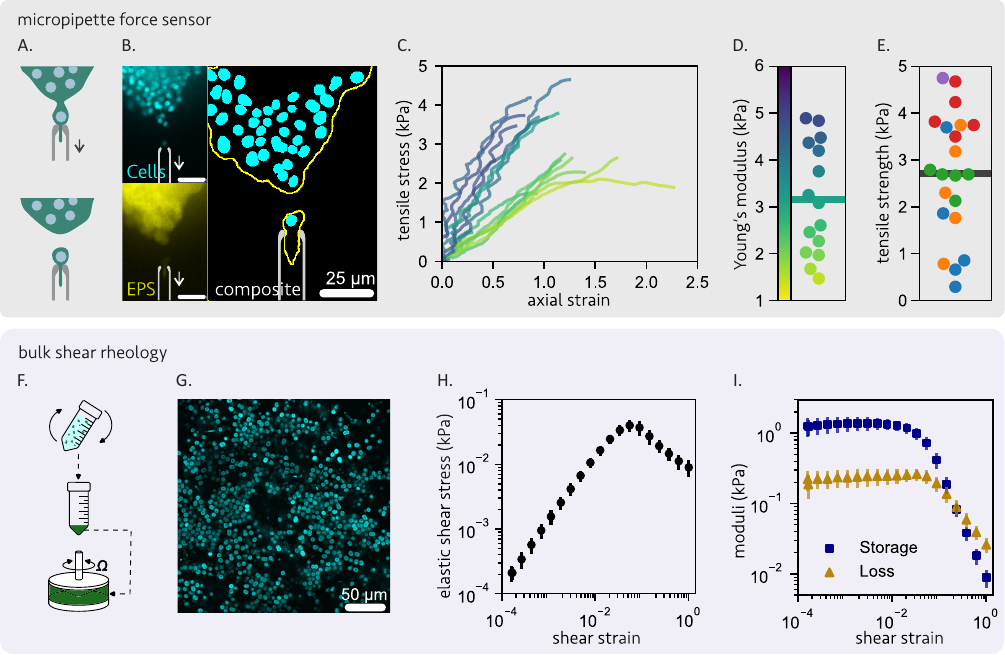}
    \caption{Mechanical properties of \textit{Microcystis} spp. colonies measured with micropipette force sensors and bulk shear rheology. (A) Schematics of the deformation of the EPS layer during a single cell pulling with a micropipette. (B) Confocal microscopy image of a colony after the removal of a single cell. Cells were visualized with chlorophyll \textit{a} auto-fluorescence, while the EPS layer was stained with 5-DTAF. Composite image indicates the masked cells and the outline of the EPS layer. The approximate position of the micropipette is illustrated. See Supplementary Movie S2. (C) Tensile stress as a function of the axial strain in the region surrounding the held cell ($N = 15$ cells from 5 colonies). (D) Young's modulus for each cell corresponding to the linear deformation regime in panel C. Horizontal bar indicates the mean value. The colormap indicates the Young's modulus for panels C and D. (E) Tensile strength corresponding to the peak tensile stress during each single-cell pulling ($N=21$ cells from 5 colonies). Horizontal bar indicates the mean value and simlar color of the symbols indicates cells pulled from the same colony. (F) Diagram of the sample preparation and measurement protocol with bulk shear rheology: A suspension of colonies was centrifuged and the supernatant was removed. The pellet formed a macroscopic aggregate containing many colonies. The aggregate was transferred to a cone-and-plate geometry, and a large-amplitude oscillatory shear was conducted ($\Omega = $ \qty{1}{rad/s}). (G) Confocal microscopy indicated the heterogeneous cell arrangement within the extracellular matrix. Cells were visualized with chlorophyll \textit{a} auto-fluorescence. (H) Elastic component of the shear stress as a function of the shear strain. (I) Storage (squares) and Loss (triangles) moduli as a function of the shear strain. Symbols and error bars indicate the mean and standard deviation, respectively, of three aggregates.}
    \label{fig:mechanics}
\end{figure}

Although the tensile force $F$ along any cross section of the colony must be uniform, the associated cross-sectional area $A_t$ was not uniform (Fig. \ref{fig:mechanics}A). Therefore the tensile stress $\sigma = {F/A_t}$ would be expected to vary along the colony. Accordingly, deformation in any region of the colony is a function of the local stress. We have conducted additional micropipette pulling experiments of colonies with stained EPS to identify the regions of largest deformation and the associated cross-sectional area (\nameref{sec:methods}). The fractured cell was encased in the EPS layer, with the cross section of the fractured area having the order of magnitude of a single cell (Fig. \ref{fig:mechanics}B). Furthermore, all cells remained spherical throughout the test, while the deformation was concentrated in the EPS layer immediately around the held cell (Supplementary Movie S2). Therefore, the tensile tests probe the mechanical properties of the two regions surrounding the two held cells (one per micropipette). We have quantified the deformation around the held cells by the axial strain $\varepsilon = \ln (1 +L_o^{-1}\,\Delta L)$ \cite{karimi_comparative_2014}, where $\Delta L$ is the increase in the distance between the micropipette tips with respect to the null-force position, and $L_o = 2\, d_1$ is the equilibrium length of the deformed region, and $d_1$ is a single cell diameter. Meanwhile, the tensile stress around the held cells, $\sigma = {F/A_1}$, was calculated based on the cross-sectional area of a single cell, $A_1 = \pi d_1^2/4$.

The trajectories of tensile stress showed a linear dependence on the axial strain followed by the fracture of the colony, which is characteristic of a linear elastic response with brittle fracture (Fig. \ref{fig:mechanics}C). However, a small fraction of the tests showed a large non-linear region, in which the colony continued to deform without fracture while the slope of the tensile stress declined. The Young's modulus $E$, defined as the slope between the tensile stress and the axial strain in the linear region, \textit{i.e.}, $\sigma = E \, \varepsilon$, had an average value of $E = 3.2 \, \pm \, 1.2$ \unit{\kPa} (mean $\pm$ SD, $N =$ 15 cells in total pulled from 5 colonies, Fig. \ref{fig:mechanics}D). Furthermore, the tensile strength $\tau_s$, defined as the maximum tensile stress until fracture, had an average value of $\tau_s = 2.7 \, \pm \, 1.3$ \unit{\kPa} ($N =$ 21 cells in total pulled from 5 colonies, Fig. \ref{fig:mechanics}E). The coefficient of variation of the tensile strength ($CV =$ 0.5) was an order of magnitude larger than the relative uncertainty of the micropipette force sensor ($\delta_{int} \, \tau_s = 0.03$) and therefore reflects heterogeneity in the microscale mechanical properties of the colonies. This mechanical heterogeneity may arise from either morphological or composition variability of the \textit{Microcystis} colonies. In the first case, variations in the area under tensile force would lead to variability in the calculated tensile strength. Although the cross-sectional area of the single cells in our samples had a large coefficient of variation ($CV =$ 0.5), there was no significant correlation observed between the average force and the average cell diameter per colony (Fig. \ref{fig:morphology}G). The second hypothesis is that mechanical heterogeneity arises from variation in EPS composition along the colony. However, we could not verify this hypothesis, and future work combining mechanical measurements with direct visualization or quantification of the chemical composition of the colonies may help resolve this question.

The microscale mechanical properties of \textit{Microcystis} spp. colonies measured with micropipette force sensors were compared with the macroscopic properties of colony aggregates measured via bulk shear rheology (\nameref{sec:methods}). \textit{Microcystis} spp. colonies were separated from the medium and concentrated into macroscopic aggregates, each containing only cells and bound EPS layer of many colonies (Fig. \ref{fig:mechanics}F). Confocal microscopy images of the colony aggregates revealed a heterogeneous structure composed of many connected colonies, without distinguishable outlines between adjacent colonies (Fig. \ref{fig:mechanics}G). The intercellular spacing was irregular and the cell volume fraction was $28 \, \pm \, 3 \%$ ($N =$ 7 frames). The rheological properties of the colony aggregates were characterized with large amplitude oscillatory shear, where the aggregates were subjected to a range of shear strains and the elastic and viscous components of the shear stress, and the associated storage and loss moduli, were measured (\nameref{sec:methods}). The elastic shear stress showed a linear dependence on the shear strain up to $\sim 10^{-2}$, indicating that the deformation is recoverable under this regime (Fig. \ref{fig:mechanics}H). Under small deformations, the aggregates had a solid-like response, characterized by a larger storage modulus than the loss modulus (Fig. \ref{fig:mechanics}I), with an average storage modulus of $G = 1.2 \,\pm \,0.3$ \unit{\kPa} ($N=$3 aggregates). At large shear strains ($\gtrsim 10^{-1}$), the macroscopic aggregates displayed a non-recoverable deformation linked to the fracture of internal structures. The transition is given by the maximum value of the elastic shear stress, defined as the yield stress $\tau_y \,=\,0.040 \pm 0.009 $ \unit{\kPa} ($N=$3 aggregates, Fig. \ref{fig:mechanics}H).

The two techniques employed here, using (i) micropipette force sensing and (ii) bulk shear rheology, probed the mechanical properties of the colonies at different length scales. While the micropipette force sensors assessed the local mechanics around single cells, the shear rheology measured the macroscopic response of a material continuum with many colonies. Moreover, the two techniques differ in the direction of deformation, with an axial deformation for the micropipette force sensing and a shear deformation for the bulk rheology. Despite these differences, some comparisons can be made between the properties measured with both techniques. In the linear elastic regime, the Young's modulus ($E$, axial) and storage modulus ($G$, shear) are related by $E = 2\,G\,(1 + \nu)$, where $\nu$ is the Poisson's ratio of the sample. Using the storage modulus $G$ measured with bulk shear rheology, we can estimate an equivalent Young's modulus of $E^* = 3.6 \,\pm \,0.9$ \unit{\kPa}, assuming a Poisson's ratio of $\nu = 0.5$ typical for polysaccharide gels \cite{subhash_concentration_2011}. The estimated value of the equivalent Young's modulus from the bulk shear rheology, $E^*$, is similar to the measured value from the micropipette force sensors, $E$. The agreement between the macroscopic and microscale properties indicates that, under small deformations, the bulk shear rheology measures the spatially-averaged elastic response of the EPS layer. In contrast, the two techniques differ substantially for large deformations. The yield stress measured with shear rheology, $\tau_y \,=\,0.040 \pm 0.009 $ kPa, was two orders of magnitude smaller than the average tensile strength measured with the micropipette pulling, $\tau_s \,= \,2.7 \, \pm \, 1.3$ \unit{\kPa} and one order of magnitude smaller than the smallest recorded tensile strength  of $\tau_s \,= \,0.3$ \unit{\kPa}. A comparison between the intercellular spacing in the colony aggregates before and after the application of shear, quantified by the average nearest-neighbour distance, did not display a significant change (ANOVA, $F(1,12) =$ 2.33, $p =$ 0.15, \nameref{sec:methods}), indicating that the individual colonies within the aggregate were not fractured during the shear rheology measurements. The low value for the yield stress and the microscopy analysis suggest that the EPS that binds the different colonies in a macroscopic aggregate yields quite easily, while the inner colonial structures are not fractured. Therefore, the shear strength of individual colonies (\textit{i.e.}, the shear stress to fracture a cell from a colony under shear deformation) is likely of the same order of magnitude as the tensile strength, although not directly probed during shear rheology. These results highlight that colonies formed by cell-division alone have stronger cell-to-cell cohesive forces (quantified here with micropipette force sensing) in comparison to the adhesive forces between aggregated colonies (quantified here with bulk shear rheology), and agree with our previous study on colony aggregation and fragmentation under shear flows \cite{Sinzato_2026}.

\subsection{Low nutrient availability leads to increased mechanical strength of \textit{Microcystis} colonies}

\begin{figure}
    \centering
    \includegraphics[]{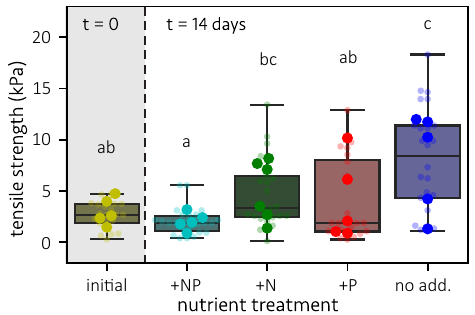}
    \caption{Tensile strength required to remove a single cell from a colony for the initial sample (t=0) and for each nutrient treatment after 14 days. Small dots indicate the tensile strength for individual cells and large dots indicate the average tensile strength of cells from the same colony. Boxes indicate the 25\textsuperscript{th}, 50\textsuperscript{th} and 75\textsuperscript{th} percentiles, while whiskers indicate the minimum and maximum values. Compact letter display indicates significantly different treatments ($p<0.05$, post-hoc Dunn's test with Benjamini/Hochberg correction, $N =$ 120 cells in total from 26 colonies).}
    \label{fig:treatment}
\end{figure}

\textit{Microcystis} spp. colonies from field samples were incubated for 14 days under different nutrient levels, namely: nutrient deplete (no add.); phosphate replete (+P); nitrate replete (+N); nutrient replete (+NP) (\nameref{sec:methods}). After incubation, the colonies were collected and the tensile strength was measured with micropipette force sensors (Fig. \ref{fig:treatment}). Similar to the initial sample (\textit{i.e.}, prior to incubation), there was a large variability in the measured tensile strength for all treatments (minimum and maximum coefficient of variation were: $CV= $0.5 and 1.0 for the initial sample and +P treatment, respectively), highlighting the mechanical heterogeneity of the colonies. As expected, the nutrient treatments had a significant effect on the tensile strength of the colonies (Kruskal-Wallis test: $H(4) = 28$, $p<0.001$). In a pairwise comparison, the median tensile strength was significantly lower for treatments containing additional P (+NP $<$ +N, +NP $<$ no add. and +P $<$ no add., p $<$ 0.05, post-hoc Dunn’s test with Benjamini/Hochberg correction). The highest median tensile strength was measured for the nutrient deplete treatment (no add.). Furthermore, only the nutrient deplete treatment had a significantly higher tensile strength than the initial sample. The results indicate that incubation under low P concentration increased the mechanical strength of \textit{Microcystis} spp. colonies, while the combination of low N and low P levels (\textit{i.e.}, no nutrient addition) further enhanced this response.

The increase in mechanical strength at low nutrient levels may be associated with changes in the EPS layer surrounding the pulled cell, as this is the region probed by the micropipette force sensors (Fig. \ref{fig:mechanics}B). Immediately prior to the force measurements, the incubated colonies were dispersed in filtered lake water, such that variations in pH and ionic strength of the medium during the measurement did not affect the tensile strength. Therefore, changes in the mechanical properties of the EPS layer must be a result of the growth of the colonies under the different nutrient conditions. A morphological analysis was performed on the measured colonies using the wide-field fluorescent images in the absence of force (Fig. \ref{fig:morphology}A). Cells within the depth of focus were segmented and the average cell diameter and average nearest-neighbour distance within a colony were calculated (Fig. \ref{fig:morphology}B). For each colony, the ratio between the combined area of the segmented cells and the area of the colony (\textit{i.e.}, the cell-to-colony area fraction) was calculated and used as a measure of the cell-to-EPS ratio (Fig. \ref{fig:morphology}C). The average tensile strength per colony was positively correlated with the average nearest-neighbour distance (Pearson correlation, $r(24)=$ 0.47, $p =$ 0.02, Fig. \ref{fig:morphology}D), while the average tensile strength was negatively correlated with the cell-to-colony area fraction (Pearson correlation, $r(24)=$ -0.45, p = 0.02, Fig. \ref{fig:morphology}E). Meanwhile, colony diameter and cell diameter did not have a significant effect on the average tensile strength per colony (Fig. \ref{fig:morphology}F and \ref{fig:morphology}G).  Previous studies have identified an increase in the EPS-to-cell ratio \cite{wang_effect_2011,duan_colony_2018,duan_ecological_2021,feng_role_2022} as well as an increase in the proportion of humic acid-like and tryptophan-like substances in the EPS fraction \cite{xiao_insights_2019,liu_excitation-emission_2017} for \textit{Microcystis} colonies cultured under low levels of N and P. The results of our morphological analysis indicate that colonies with a higher EPS-to-cell ratio have greater mechanical strength. However, it is not clear whether this increase in colony strength was due solely to differences in proportion of EPS to cells, or if the composition of the EPS could also have played a role. A correlation between the mechanical properties of the EPS fraction and its composition was identified in bacterial biofilms \cite{ohmura_vivo_2024,savorana_stress-hardening_2025}, and thus may also be relevant in cyanobacterial colonies.

\subsection{Mechanical properties of \textit{Microcystis} colonies in comparison with similar biofilms}

\begin{table}[t!]
    \centering
    \caption{Published measurements of mechanical properties of (cyano)bacterial biofilms, aggregates and colonies}
    \label{tab:mech_prop}
    \footnotesize
    \begin{tabular}{lllll}
         \textbf{Sample} & \textbf{Method} & \textbf{Property measured} & \textbf{Results} & \textbf{Reference}\\
         \hline
         Bacterial biofilm  & Micropipette force sensor & Tensile strength & 0.06 to 19 \unit{\kPa} & \cite{aggarwal_development_2010} \\ %(\textit{P. aeruginosa}) Intact
         Bacterial biofilm  & Micropipette force sensor & Tensile strength & 0.4 to 15 \unit{\kPa} & \cite{poppele_micro-cantilever_2003} \\ % (\textit{P. aeruginosa}) Detached
         Mixed biofilm & Micropipette force sensor & Tensile strength & 2 to 55 \unit{\kPa} & \cite{aydin_strong_2026} \\ % Bacterial and fungal 
         Cyanobacterial colonies &  Micropipette force sensor & Tensile strength & 0.3 to 18 \unit{\kPa} & This study \\ % 
         Cyanobacterial mat &  Bulk tensile test & Tensile strength & 1 to 20 \unit{\kPa} & \cite{vignaga_quantifying_2012} \\ % (\textit{Phormidium} sp.)
         Bacterial biofilm &  Bulk tensile test & Tensile strength & 0.5 to 1 \unit{\kPa} & \cite{ohashi_novel_1999} \\
         Bacterial biofilm & Thin-film cantilever & Tensile strength & 0.1 to 0.6 \unit{\kPa} & \cite{croland_strengthening_2025} \\ % (\textit{B. subtilis}) 
         Cyanobacterial colonies &  Cone-and-plate shear & Erosion stress & 4.5$\cdot$10\textsuperscript{-5} to 1.9$\cdot$10\textsuperscript{-3} \unit{\kPa} & \cite{Sinzato_2026} \\ % (\textit{M.} strain V163) 
         Cyanobacterial aggregate &  Bulk shear rheology & Yield stress & 4.3$\cdot$10\textsuperscript{-3} $\pm$ 0.3$\cdot$10\textsuperscript{-3} \unit{\kPa} & \cite{Sinzato_2026} \\ % (\textit{M.} strain V163) 
         Cyanobacterial aggregate &  Bulk shear rheology & Yield stress & $0.040 \pm 0.009 $ \unit{\kPa} & This study \\
         Bacterial biofilm &  Bulk shear rheology & Yield stress & 0.023 ± 0.005 \unit{\kPa} & \cite{pavlovsky_effects_2015} \\ % (\textit{S. epidermidis})
         Mixed biofilm &  Bulk shear rheology & Yield stress & 5$\cdot$10\textsuperscript{-4} to 1$\cdot$10\textsuperscript{-3} \unit{\kPa} & \cite{towler_viscoelastic_2003} % (\textit{S. epidermidis})
         
    \end{tabular}
\end{table}

To the best of our knowledge, this is the first study that measured the tensile strength of natural colonies of \textit{Microcystis}. In our previous work on laboratory cultures of \textit{Microcystis}, we identified an erosion stress of 4.5$\cdot$10\textsuperscript{-5} to 1.9$\cdot$10\textsuperscript{-3} \unit{\kPa} required to erode cells from a colony under a cone-and-plate flow \cite{Sinzato_2026}. In the same study, we identified a yield stress of 4.3$\cdot$10\textsuperscript{-3} $\pm$ 0.3$\cdot$10\textsuperscript{-3} \unit{\kPa} for a colony aggregate under shear rheology. The yield stress and erosion stress measured for these laboratory cultures of \textit{Microcystis} were much lower than the yield stress and tensile strength, respectively, measured in this study for natural samples of \textit{Microcystis}. Morphological differences are commonly observed between colonies from natural samples and laboratory cultures of \textit{Microcystis} \cite{xiao_colony_2018} and may be connected to the differences in mechanical strength \cite{Sinzato_2026}. Extending the analysis to recent studies on (cyano)bacterial biofilms, a wide range of values has been reported according to the technique and species investigated (Table \ref{tab:mech_prop}). The tensile strength of cyanobacterial mats of \textit{Phormidium} sp. measured by bulk tensile test \cite{vignaga_quantifying_2012} was similar to our results for \textit{Microcystis}. Previous studies with micropipette force sensors on mono and mixed-species heterotrophic bacterial biofilms reported a large variability in the tensile strength \cite{aggarwal_development_2010,poppele_micro-cantilever_2003,aydin_strong_2026}. In contrast, bulk tensile tests identified a smaller variability between samples \cite{ohashi_novel_1999,croland_strengthening_2025}, with values comparable to the lower end of the range measured with micropipette force sensors. These results indicate that the local mechanical properties of bacterial biofilms are highly heterogeneous, while bulk properties are limited by the weakest regions. When comparing solely bulk properties, measurements of yield stress of bacterial biofilms under bulk shear rheology \cite{towler_viscoelastic_2003,pavlovsky_effects_2015} reported much lower failure stress levels than those measured by bulk tensile tests \cite{ohashi_novel_1999,croland_strengthening_2025}. Therefore, the mode of application of stress (traction \textit{vs} shear) has a large impact on the mechanical response of biofilms.

In the context of the environmental forces experienced by \textit{Microcystis} colonies, the ecological advantage of strong mechanical properties remains unclear. Even the lowest measured value for the tensile strength is many orders of magnitude higher than the hydrodynamic stresses produced by natural wind mixing ($\tau \sim$ 10\textsuperscript{-4} to 10\textsuperscript{-2} \unit{Pa}) \cite{wang_field_2016} or artificial mixing systems ($\tau \sim$ 10\textsuperscript{-2} to 10\textsuperscript{-1} \unit{Pa}) \cite{visser_artificial_2016,lai_turbulent_2019}. This means that, in line with our previous results \cite{Sinzato_2026}, all colonies in our study would resist flow-induced fragmentation, with no clear advantage between nutrient replete and nutrient deplete samples. In addition to hydrodynamic stress alone, biotic interactions such as grazing may impose strong selective pressures on colony cohesion. The EPS layer of \textit{Microcystis} colonies has been proposed as a key defense against grazing and digestion \cite{reynolds_variability_2007}. Comparative studies have shown that \textit{M. aeruginosa} colonies can be strongly grazed by amoebae, whereas \textit{M. viridis} colonies with a thicker EPS layer are more resistant \cite{van_wichelen_strong_2010}. Grazing pressure itself can also shape colony morphology: flagellate grazing induces colony formation in \textit{M. aeruginosa} \cite{yang_morphological_2006}, and similarly, filter-feeders impose selection for multicellularity in green algae \cite{herron_novo_2019}. However, the protective role of EPS thickness remains debated, as other studies reported no correlation between EPS thickness and grazing efficiency, suggesting that EPS composition may be more important \cite{van_wichelen_importance_2012}. Beyond chemical protection, high tensile strength in \textit{Microcystis} colonies may mechanically inhibit fragmentation into ingestible units under predator-imposed forces, a mechanism analogous to the resistance of bacterial biofilms to phagocytosis \cite{stewart_biophysics_2014}.

\section{Conclusions}

This study presented direct measurements of the local and bulk mechanical properties of natural cyanobacterial colonies, focusing on samples of \textit{Microcystis} spp. Simultaneous micropipette force sensing and fluorescence microscopy revealed that colony tensile strength arises primarily from the bound EPS layer surrounding individual cells, with pronounced mechanical heterogeneity both within and between colonies. Comparison with bulk shear rheology further showed that the measured mechanical properties depended strongly on the mode and length scale of the applied stress. Notably, our results showed that the stresses required to yield or fracture colonies exceed typical hydrodynamic stresses in lakes by several orders of magnitude, leaving open whether strong mechanical resistance is an adaptive trait (potentially selected through biotic interactions such as grazing) or an emergent consequence of colony formation itself. Furthermore, nutrient availability significantly altered colony tensile strength, motivating future work that explicitly links mechanical properties to colony morphology and EPS metabolism. Beyond advancing a mechanical understanding of cyanobacterial colonies, our results may help inform the design of mitigation strategies to prevent toxic cyanobacterial blooms \cite{visser_artificial_2016} and are likely to be relevant to other applied contexts involving microbial aggregates.

\section{Methods\label{sec:methods}}

\subsection{Micropipette force sensors\label{sec:met_force_sensors}}

Micropipette force sensors were used to measure the mechanical properties of \textit{Microcystis} spp. colonies, in which the tensile force  $F$ applied to the held colony was calculated from the deflection $\Delta y$ of a flexible micropipette relative to its rest position, such that $F = k_p \,\Delta y $, where $k_p$ is the bending stiffness of the micropipette. \cite{backholm_micropipette_2019}. Glass micropipettes were fabricated in a micropipette puller (P-1000, Sutter Instrument) and shaped to a 90-degree bend with a microforge (MF2, Narishige). One flexible micropipette was used for the measurements on the initial sample (t=0) and another micropipette was used for the samples after the incubation ($t=$ 14 days), each with a bending stiffness of $k_b=0.63 \pm 0.02$\,\unit{nN/\um} (mean $\pm$ SD, $N=$ 3 calibration tests) and $3.47 \pm 0.04$\,\unit{nN/\um} ($N=$ 3) respectively. Each flexible micropipette was calibrated by measuring its relative deflection against a pre-calibrated micropipette (Fig. \ref{fig:pipette_calibration}B), which in turn was calibrated with the hanging droplet method (Fig. \ref{fig:pipette_calibration}A) \cite{backholm_micropipette_2019}.

\textit{Microcystis} spp. colonies tested with the micropipette force sensing were collected from Lake Braassemermeer, the Netherlands, on 12 August 2024 (52°11'47"N 4°38'37"E). Lake water was sieved through a \qty{200}{\um} net to remove large zooplankton and transported to the laboratory. A subsample of colonies was disaggregated in sucrose solution (7 \unit{g/L}), and the single cell diameter was measured ($d_1 = 4.6 \,\pm\, 1.3$ \, \unit{\um}, mean $\pm$ SD, $N =$ 10285 cells). Before each force measurement, the suspension of colonies was sieved through a \qty{200}{\um} net to remove large colonies and then concentrated at a \qty{35}{\um} net. The collected colonies were re-dispersed in filtered (\qty{0.2}{\um}) lake water. The size range of 35 to 200\,\unit{\um} was selected to exclude small colonies with too few cells and large colonies which did not fit in the field of view of the imaging system. The suspension of colonies was gently injected in a chamber with a gap of \qty{2}{mm} and two lateral openings for access to the micropipettes. The laboratory was kept at \qty{20}{\degreeCelsius}. A colony was first aspirated with a rigid micropipette and brought near the flexible micropipette, which aspirated the colony at the opposite side. Next, the colony was pulled by the rigid pipette at a constant speed of \qty{1}{\um/s}. The inner diameter of the micropipettes was smaller than the diameter of a single cell such that each micropipette held the colony by one cell only. Preliminary tests with micropipettes with diameters larger than a single cell led to complete aspiration of cells into the tubing instead of a successful hold, due to the large suction pressure required. The deflection of the flexible micropipette was tracked until the full fracture of the colony, identified by the return of the flexible micropipette to its rest position. The translation of the micropipettes was controlled with a pair of automated micromanipulators (uMp3, Sensapex, Finland). Phase-contrast microscopy was used to image each micropipette, while wide-field fluorescence microscopy with emission from chlorophyll \textit{a} (excitation at 604-644 \unit{\nm}) was used to image the colony. Tensile tests without a fragment due to loss of suction pressure were discarded (18\% of tests). For a subset of the pulled cells in the initial sample (6 out of 21 cells), the rest length of colony and the axial strain $\varepsilon$ could not be calculated, as the tensile test started from a pre-stretched position.

\subsection{Image analysis}

The time-lapse videos of tensile tests were pre-processed with Gaussian blur and background subtraction. The translation over time $y(t)$ of the micropipettes was detected with template matching, using the 90-degree bend as the reference point. The morphology of each colony tested was analysed with an image processing script based on the python library \textit{scikit-image}. A wide-field fluorescent image of the colony in a rest state (no applied force) was captured. The colony regions within focus were masked with a top-hat filter and Otsu threshold. Cells within focus were segmented and measured with Voronoi-Otsu labelling. The average nearest-neighbour Distance (NND) between the cells was computed for each colony. The colony envelope (combined cells and EPS-filled gaps) was masked with morphological closing, and its area and equivalent circular diameter were computed. The cell-to-colony area fraction was computed as the ratio between the total area of the labelled cells and the total area of the colony envelope.

\subsection{Visualization of EPS layer}

\textit{Microcystis} spp. colonies used for visualization of EPS layer were collected from Lake Haarlemmermeerse Bosplas, the Netherlands, on 10 October 2025. Buoyant colonies were separated from the sinking fraction and a pressure of \qty{1}{\MPa} was applied to collapse gas vesicles. The colonies were stained for 24 hours with 1 \unit{mM} of 5-(4,6-Dichlorotriazinyl)aminofluorescein (5-DTAF, AAT Bioquest) in a carbonate buffer (0.1 \unit{M}, pH 9). Excess stain was washed off four times with carbonate buffer. The suspension of colonies was injected into a chamber, and cells were pulled with two rigid micropipettes. Dual-channel time stacks were captured in a spinning disk confocal microscope. The cells were visualized with the chlorophyll \textit{a} fluorescence (excitation at 640 \unit{nm}) and the EPS layer was visualized with the 5-DTAF fluorescence (excitation at 470 \unit{nm}). Cells were segmented using Stardist \cite{schmidt_cell_2018}. The outline of the EPS layer was segmented with Gaussian blur, background subtraction, and triangle threshold.

\subsection{Bulk shear rheology of macroscopic colony aggregates}

The bulk mechanical properties of macroscopic aggregates of concentrated \textit{Microcystis} spp. colonies were measured with shear rheology. \textit{Microcystis} spp. colonies were collected from Lake Braassemermeer on 18 July 2025. The plankton sample was left to sediment for 24 hours at \qty{4}{\degreeCelsius} and the buoyant fraction was collected. The suspension was subsequently pressurized at \qty{1}{\MPa} to collapse the gas vesicles. The colonies were separated from the medium following the protocol described by Xu \textit{et al.} \cite{xu_investigation_2013}. The suspension was centrifuged at 2500 g, \qty{4}{\degreeCelsius} for 15 minutes. The supernatant containing the soluble EPS was discarded and the pellet containing the cells embedded in the bound EPS (combined loosely bound and tightly bound) formed a macroscopic aggregate. The aggregates were prepared in three replicates and were separated for the rheological test. For each aggregate, a volume of $\approx$ \qty{0.6}{mL} was loaded into a cone-and-plate rheometer equipped with a humidity chamber to reduce evaporation. The upper conical probe had \qty{50}{mm} diameter, \ang{1} cone angle, and a sand-blasted surface. The bottom plate was covered with sandpaper to prevent wall slip. After loading the aggregate, a pre-shear of \qty{100}{s^{-1}} was applied for \qty{30}{s}, followed by \qty{300}{s} of rest. Large amplitude oscillatory shear (LAOS) at $\omega =$ \qty{1}{rad/s} and \qty{20}{\degreeCelsius} was conducted in the aggregate. LAOS consists of an oscillatory shear with a shear strain of amplitude $\gamma_o$, where the in-phase (or out-of-phase) component of the shear stress is the elastic (or viscous) stress, denoted as $\tau_e\,(\gamma_o)$ (or $\tau_v\,(\gamma_o)$). The ratio between the elastic (or viscous) stress and the shear strain amplitude is the storage (or loss) modulus, denoted as $G'\,(\gamma_o)$ (or $G''\,(\gamma_o)$). We denote $G$ as the limit of the storage modulus for small deformations, $G = G'\,(\gamma_o \rightarrow 0)$. Measurements were repeated for three aggregates. The structure of one colony aggregate, before and after the application of shear, was imaged in a spinning disk confocal microscope. The cells were visualized with the chlorophyll \textit{a} fluorescence (excitation at 640 \unit{nm}) and segmented using Stardist \cite{schmidt_cell_2018}. The volume fraction of cells within the aggregate was computed from the cell area fraction measured in the confocal vertical slices. The average nearest-neighbour distance (NND) was computed for 7 frames, and the mean values before and after the application of shear were compared with a one-way ANOVA test.

\subsection{Nutrient treatments}

Natural phytoplankton communities were sampled from Lake Braassemermeer on 12 August 2024 and filtered over a \qty{200}{\um} sieve to remove larger zooplankton. The water was subsequently transferred to \qty{10}{L} flasks and incubations were performed at \qty{22}{\degreeCelsius} with an incident irradiance of \qty{30}{\micro\mole.photons.m^{-2}.s^{-1}} at a light-dark cycle of 15h:9h. Cultures were exposed to four nutrient treatments (1 flask per treatment) that included nutrient deplete (No add.), phosphate replete (+P), nitrate replete (+N), and nutrient replete (+NP) treatments. The total nitrogen (TN) and total phosphorus (TP) concentrations in the initial sample were TN = \qty{1.12}{mg/L} and TP = \qty{0.20}{mg/L}, and the concentrations were raised to TN = \qty{10.83}{mg/L} and TP = \qty{1.21}{mg/L} by adding either NaNO\textsubscript{3} or K\textsubscript{2}HPO\textsubscript{4} to the corresponding nutrient treatment. Temperature was regulated with a climate control system (SpecView 32/859, SpecView Ltd., Uckfield, UK). At least 5 colonies were collected from each nutrient treatment after 14 days and tested with micropipette force sensing following the procedures described in section \ref{sec:methods}.\ref{sec:met_force_sensors}.

\section*{Acknowledgments}

We are grateful to Pieter Slot, Merijn Schuurmans, and Axel Gunderson for support with fieldwork, Qianwei Li, Dennis Waasdorp and Suzanne Wiezer for assistance with fieldwork and laboratory measurements, Erik Hop for support with the experimental setup design, Eva Pot for assistance with the setup calibration, Nico Schramma for support with image processing. We especially thank Johan Oosterbaan and Richard Steel for their contributions during the project's conception. We acknowledge funding from the Hoogheemraadschap van Rijnland (Rijnland Regional Water Authority). MJ acknowledges support from the ERC grant no.~"2023-StG-101117025, FluMAB. AMD and DBVdW are funded by the European Union (ERC, BLOOMTOX, project number 101044452). Views and opinions expressed are, however, those of the author(s) only and do not necessarily reflect those of the European Union or the European Research Council Executive Agency. Neither the European Union nor the granting authority can be held responsible for them.

\section*{Data and code availability}

Primary experimental data and scripts for image analysis are available at \url{https://doi.org/10.21942/uva.32064591.v1}.

\printbibliography

\end{refsection}

%%%%%%%%%% 
% SUPPORTING INFORMATION
%%%%%%%%%%%

%Supporting Information
% Clear page and reset figure,equation and page counter
\clearpage
\setcounter{page}{1}
\renewcommand{\thefigure}{S\arabic{figure}}
\setcounter{figure}{0}
\renewcommand{\theequation}{S\arabic{equation}}
\setcounter{equation}{0}

% This adds line numbers to the manuscript
%\linenumbers

\begin{refsection}

\section*{Supporting Information for}

\textbf{Title: } On Linear and non-Linear Mechanics of Cyanobacterial Colonies
\\
\textbf{Authors: }Yuri Z. Sinzato, Annemieke M. Drost, Dedmer B. Van de Waal, Robert Uittenbogaard, Petra M. Visser, Jef Huisman and Maziyar Jalaal
\\
\textbf{Corresponding authors: }Maziyar Jalaal (m.jalaal@uva.nl, m.jalaal@damtp.cam.ac.uk) / Yuri Z. Sinzato (y.z.sinzato@uva.nl)

% Table of contents for the SI
\subsection*{Contents: }
\vspace{0.2cm}
\renewcommand{\listfigurename}{Supplementary Figures}
%\localtableofcontents
\locallistoffigures

\clearpage

%\subsection*{Supporting Information Text}
%\printbibliography
%\clearpage

\subsection*{Supplementary Figures}

\begin{figure}[h!]
    \centering
    \includegraphics[]{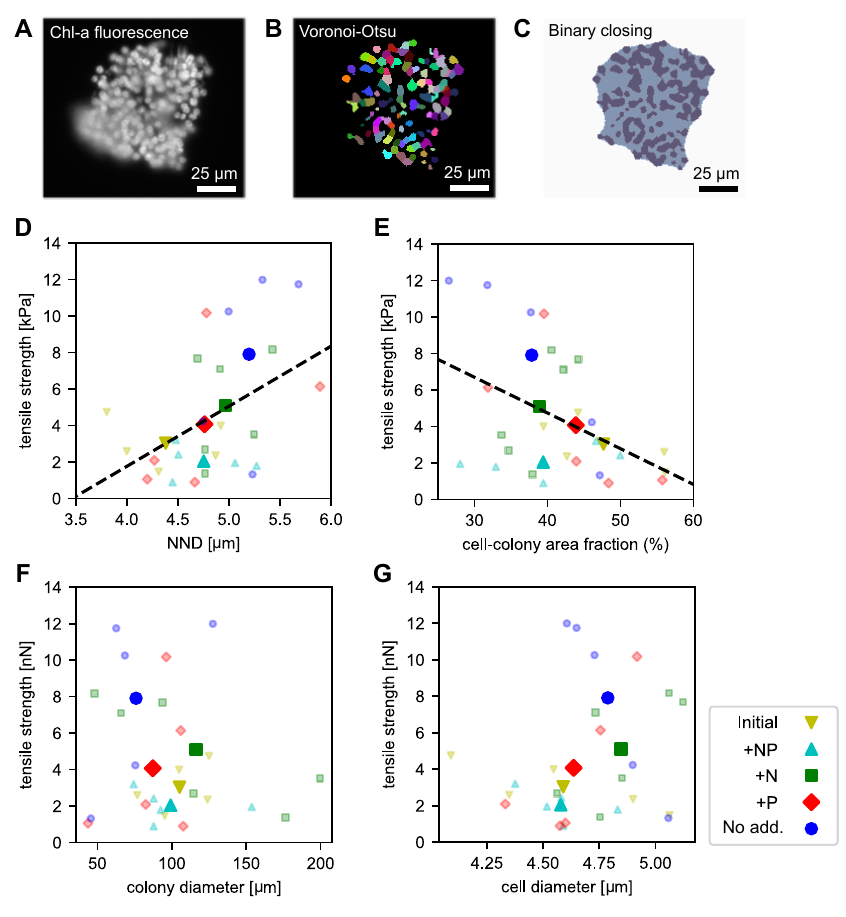}
    \caption[Relation between mechanical properties and morphology of colonies]{Relation between mechanical properties and morphology of colonies. (A) Wide-field fluorescence imaging of cells using the excitation of chlorophyll \textit{a}. (B) Cells within focus are segmented using a Voronoi-Otsu method. (C) Colony envelope is obtained using binary closing operation. (D) Average tensile strength increases with the average nearest-neighbor distance (NND) (Pearson correlation, $r(24)=$ 0.47, $p =$ 0.02). (E) Conversely, the average tensile strength of the colonies decreases with the area fraction between cells and the colony (Pearson correlation, $r(24)=$ -0.45, $p =$ 0.02). (F) No significant correlation was found between the average tensile strength and colony diameter (Pearson correlation, $r(24) =$ -0.16, $p=$ 0.43), (G) as well as with the cell diameter (Pearson correlation, $r(24) =$ 0.17, $p=$ 0.40). Small symbols indicate the average tensile strength of individual colonies, while large symbols indicate the average of treatments. Number of colonies = 26}
    \label{fig:morphology}
\end{figure}

\clearpage

\begin{figure}[h!]
    \centering
    \includegraphics{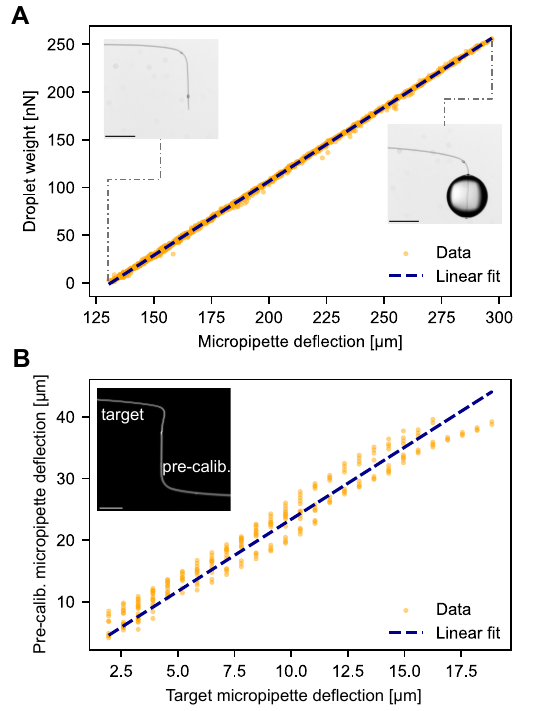}
    \caption[Micropipette calibration]{Calibration curves of flexible micropipettes. (A) Calibration of a reference micropipette via the hanging water droplet method. Weight of water droplet is measured as a function of the deflection of the micropipette (symbols) and the data is fitted with a linear regression: $F_d = k_b \, \Delta y$, where $k_b = $ 1.550 $\pm$ $0.001$\,\unit{nN/\um} is the bending stiffness. Insets depict the droplet at the minimum and maximum volumes. (B) Calibration of a target micropipette via the relative stiffness method. The deflection of a pre-calibrated micropipette (symbols) is measured as a function of the deflection of a target micropipette and the data is fitted with a linear regression: $\Delta y_{p} = a \, \Delta y_t$, where $a = k_p/\,k_t = $ 2.15 $\pm$ $0.03$ is the ratio of bending stiffnesses. Inset depicts a phase-contrast image of the micropipettes. Scale bar = \qty{200}{\um}}
    \label{fig:pipette_calibration}
\end{figure}

\clearpage
Supplementary movies are available in the online version:
\begin{itemize}
    \item Supplementary Movie S1 - Micropipette force sensing test on a \textit{Microcystis} colony. Top micropipette was flexible, while the bottom micropipette was rigid and translated at \qty{1}{\um/s}. Frames are composite images, where the micropipettes were visualized with phase contrast (magenta) and cells were visualized with wide-field fluorescence (cyan). Video was sped up 40x times.
    \item Supplementary Movie S2 - Visualization of stained EPS layer of a \textit{Microcystis} colony during pulling. Micropipette was rigid and translated at \qty{1}{\um/s}. Frames are a montage of each confocal microscopy channel (EPS in yellow and cells in cyan) and a composite image indicating the segmented cells (cyan) and the outline of the EPS layer (yellow). Video was sped up 10x times.
\end{itemize}

\end{refsection}

\end{document}